\documentclass[epj]{svjour}

\usepackage{graphicx}
\usepackage{fancyhdr}
\def\makeheadbox{{
}}
\def\mytitle{Outlook from SUSY07} 
\def\myauthors{John Ellis}  
\def\mytype{Plenary}
\def\mysession{John Ellis}

\begin{document}
\title{Outlook from SUSY07}
\author{John Ellis\inst{1}
\thanks{\emph{Email:} John.Ellis@cern.ch}%
}                     
%
%
\institute{Theory Division, Physics Department, CERN, CH 1211 Geneva 23, Switzerland}
%
\date{}
%
\abstract{
\vskip - 2in
\rightline{CERN-PH-TH/2007-204}
\rightline{October 2007}
\vskip 1.5in
Make-or-break time is near for the Higgs boson and supersymmetry. The
LHC will soon put to the sword many theoretical ideas,
and define the future for collider physics.
} 
\maketitle
%

\section{Introduction}
\label{intro}

I was in Karlsruhe in 1969 when the first astronauts landed on the Moon,
realizing President Kennedy's commitment: `this nation should commit itself to 
achieving the goal ... of landing a man on the Moon and returning him safely to the Earth'.
Likewise, in 1994 the CERN Council committed the Organization to the goal of
discovering the Higgs boson and supersymmetry (if they exist) at the LHC. Now, back in
Karlsruhe, we supersymmetrists face the exhilarating prospect that our cherished ideas
will soon be subjected to the ordeal of experimental test~\cite{Wilczek}.

The organizers of this meeting explicitly exonerated me from giving a summary
talk, asking me instead to present a personal outlook on the future (as offered by the LHC {\it et al}).
Nevertheless, I have not quite taken them at their words, and base (at least
some of) my talk on presentations made at SUSY07.

\section{Hunting for the Higgs Boson}
\label{Higgs}

The LHC, with its centre-of-mass energy of 14 TeV and its nominal luminosity
of $10^{34}$~cm$^{-2}$s$^{-1}$ and the possibility of an upgrade to
$10^{35}$~cm$^{-2}$s$^{-1}$~\cite{Evans}, offers the best prospects for discovering
new physics beyond the Standard Model (SM). 
Since interesting cross sections such as those for
supersymmetry and the Higgs boson are typically ${\cal O}(1)$/(1 TeV)$^{2}$,
possibly with small prefactors ${\cal O}(\alpha^2)$,
whereas the total cross section is ${\cal O}(1)$/(100~MeV)$^{2}$, looking for
this interesting new phys- ics will be like looking for a needle in 100,000 haystacks.

\begin{figure}
\includegraphics[width=0.5\textwidth,height=0.35\textwidth,angle=0]{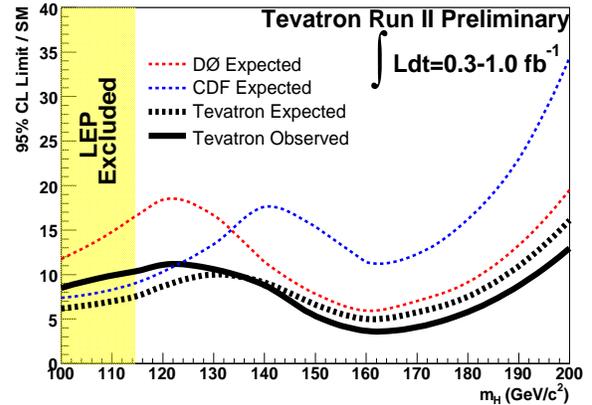}
\caption{Current combined upper limit from CDF and D0 on the production of a SM Higgs boson at the 
Tevatron~\protect\cite{Duperrin}.}
\label{fig:1}       
\end{figure}

The Tevatron may have a chance of pipping the LHC 
in the race for the Higgs boson. Already, as seen in Fig.~\ref{fig:1}, the
sensitivity of the combined CDF and D0 searches is within an order of
magnitude of the cross section expected for the SM Higgs boson
from the LEP lower limit up to $m_H \sim 190$~GeV, the mass range allowed at the 95~\%
confidence level~\cite{Duperrin}, and the sensitivity is within a factor 2 of the SM for
$m_H \sim 160$~GeV. Soon we may know whether the Higgs is either close to
the LEP limit or in an `unlikely' range.

\begin{figure*}
\includegraphics[width=1.\textwidth,height=0.5\textwidth,angle=0]{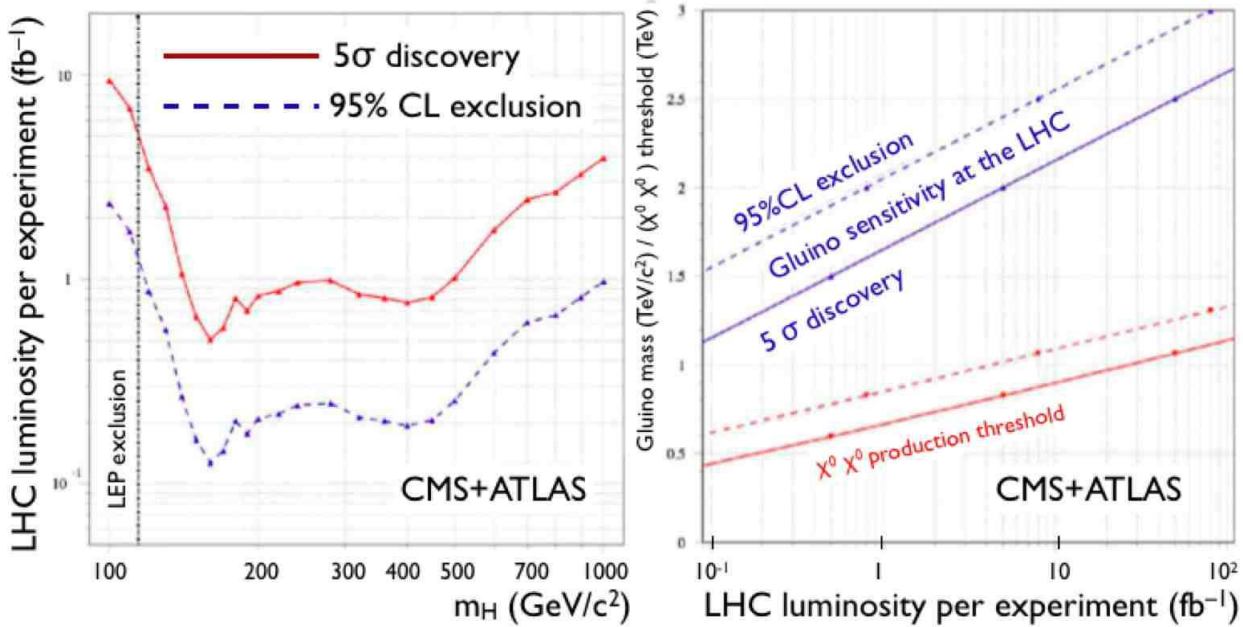}
\caption{The combined sensitivities of ATLAS and CMS to a
Standard Model Higgs boson (left), and the gluino (right), as a function of the
analyzed LHC luminosity. The right panel also shows the threshold for
sparticle pair production at a LC for the corresponding gluino mass, calculated
within the CMSSM~\protect\cite{POFPA}.}
\label{fig:POFPA}       
\end{figure*}

The search for the Higgs boson at the LHC will require combining various
different signatures~\cite{Jakobs}, including $\gamma \gamma$, four-lepton final states,
$\tau \tau$, ${\bar b} b$, $WW$ and $ZZ$. As seen in in Fig.~\ref{fig:POFPA},
combining searches by ATLAS
and CMS, 200~pb$^{-1}$ should suffice to exclude a SM between about 140
and 500~GeV, 1~fb$^{-1}$ should enable a SM Higgs boson to be discovered
with 5-$\sigma$ significance over a similar mass range, and 5~fb$^{-1}$ 
should enable a discovery discover whatever its mass~\cite{POFPA}.
Eventually, if the Higgs mass $\sim 120$~GeV,
it should be possible at the LHC to measure SM Higgs couplings to
$\tau \tau$, ${\bar b} b$, $WW$ and $ZZ$ with an accuracy $\sim 20$~\%, and
there are also prospects for measuring the Higgs spin via its decays into $ZZ$~\cite{LHCH}.

One of the biggest puzzles in Higgs physics is its contribution to vacuum
energy. The naive Higgs potential $- \mu^2 |H|^2 + \lambda |H|^4$ makes a
negative contribution to the vacuum energy that is negative and some 60 orders of
magnitude larger than the physical value of the dark energy. Some mysterious
mechanism is needed to cancel this Higgs contribution to 60 decimal places.
What are we missing? Are we barking up the wrong tree?

\section{Why Supersymmetry?}
\label{why}

There are many motivations for supersymmetry, including its
intrinsic beauty, its help in rendering the hierarchy of mass scales in
fundamental physics more natural, its help in unifying the gauge couplings,
its prediction that the Higgs boson should be relatively light: $m_H < 150$~GeV
as suggested by precision electroweak data, and its offer of a natural cold dark matter
candidate~\cite{EHNOS}. Moreover, it is (almost) an essential ingredient in string theory.

There have recently been several impressive pieces of direct observational evidence
for collisionless cold dark matter, e.g., the `bullet cluster' which has been shown by
weak lensing to contain two lumps of dark matter that have passed through each other,
while the the associated gas clouds have collided, heated up and remained stuck in
between~\cite{Clowe}. On the other hand, there are problems for the cold dark matter paradigm
provided, e.g., by dwarf spheroidal galaxies~\cite{Gilmore}, 
the abundances of satellites of the Milky 
Way, and the apparent absence of cusps in galactic centres. Are we barking up the wrong
tree again?

\section{Constraints on Supersymmetry}
\label{what}

There are important direct constraints on supersymmetry due to the
absence of sparticles at LEP and the Tevatron, and also indirect constraints from, e.g.,
the LEP lower limit of 114~GeV on the Higgs mass, the success of SM
calculations of $b \to s \gamma$, etc. One of the most important constraints is that
imposed by the cold dark matter density, assuming it is largely composed of the
lightest supersymmetric particle (LSP): $0.094 < \Omega_{LSP} h^2 < 0.124$. There
is still some debate about the interpretation of the BNL measurement of the
anomalous magnetic moment of the muon ($g_\mu - 2$), which now disagrees by 3.4 $\sigma$
with a SM calculation based on low-energy $e^+ e^-$ data~\cite{Czarnecki}. Recent $e^+ e^-$
data agree very well with earlier data, whereas preliminary new $\tau$ decay data
apparently disagree with previous data.

Presumably the LSP has no strong or electromagnetic interactions, otherwise 
it would bind to conventional matter and be detectable as anomalous heavy nuclei.
Possible weakly-interacting scandidates include the sneutrino (though this seems
to be excluded by LEP and direct searches), the lightest neutralino $\chi$ 
(a mixture of the spartners of the $Z, H$ and $\gamma$), and the gravitino
(which would be a nightmare for astrophysical detection, but a boon for colliders,
as discussed later).

\section{A Paradigm: the CMSSM with a neutralino LSP}
\label{CMSSM}

For the rest of this talk, I focus on the minimal supersymmetric extension of 
the Standard Model (MSSM), including two Higgs doublets with coupling $\mu$
and a ratio of  v.e.v.Õs denoted by $\tan \beta$. The MSSM has {\it a priori}
unknown supersymmetry-breaking parameters, scalar masses $m_0$, gaugino 
masses $m_{1/2}$, trilinear soft couplings $A_0$, and a bilinear soft coupling $B_0$.
Uuniversality at the input GUT scale is often assumed, the constrained MSSM (CMSSM)
framework with a single $m_0$, a single $m_{1/2}$, and a single $A_0$. However, 
there is no necessity for the universality hypothesis in string theory. I emphasize that
{\it the CMSSM is not the same as minimal supergravity} (mSUGRA), which imposes an
relation on the gravitino mass: $m_{3/2} = m_0$, and the additional relation
$B_0 = A_0 Ð m_0$.

\begin{figure}
\resizebox{0.95\columnwidth}{!}
{\includegraphics{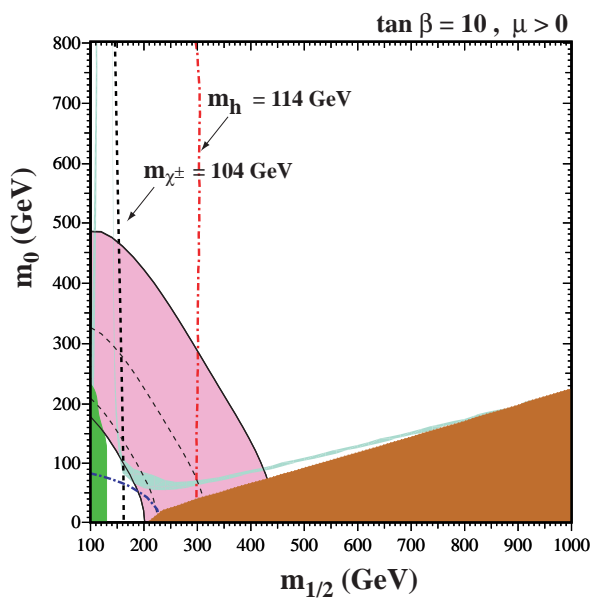}}
\caption{\label{fig:CMSSM} The $(m_{1/2}, m_0)$ plane in the CMSSM for
$\tan \beta = 10$, $\mu > 0$ and $A_0 = 0$~\protect\cite{EOSS}, 
incorporating the theoretical, experimental
and cosmological constraints described in the text.}
\end{figure}

\begin{figure}
\resizebox{1.03\columnwidth}{!}
{\includegraphics{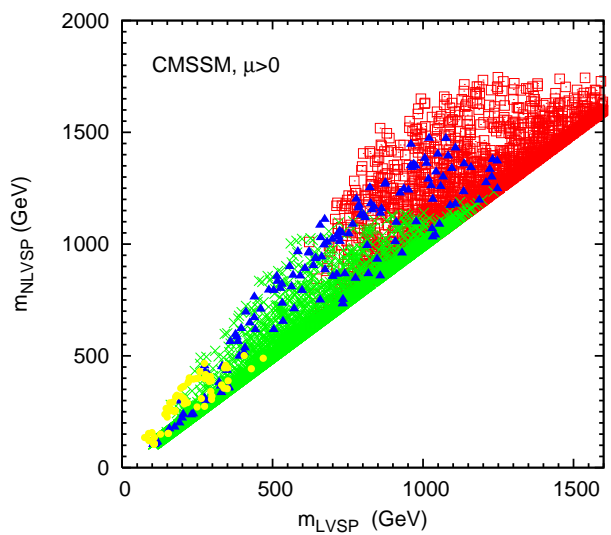}}
\caption{\label{fig:NLSP} The masses of the
lightest and next-to-lightest visible supersymmetric particles in a sampling
of CMSSM scenarios~\protect\cite{NLSP}. Also indicated are the scenarios
providing a suitable amount of cold dark matter (blue), those detectable at the LHC
(green) and those where the astrophysical dark matter might be detected
directly (yellow).}
\end{figure}

Fig.~\ref{fig:CMSSM} is an example of the current constraints on the CMSSM~\cite{EOSS},
assuming that the LSP is the lightest neutralino $\chi$, showing the region
excluded because the LSP is the charged stau (shaded brown), excluded by
$b \to s \gamma$ (shaded green), preferred by $g_\mu - 2$ (shaded pink)
and by the cold dark matter density (shaded pale blue). The region allowed
by these constraints extends to large $m_{1/2}$ (and there is another
allowed region at large $m_0$), so sparticles may be quite heavy, as seen
in Fig.~\ref{fig:NLSP}~\cite{NLSP}. The red symbols are the full data sample, the blue
symbols indicate models that could provide the astrophysical dark matter,
the green symbols indicate models that are detectable at the LHC, and the
yellow points are likely to be detectable directly in searches for dark matter
scattering~\cite{Hooper}.

\section{What Might be the Scale of Supersymmetry?}
\label{where}

\begin{figure*}[tbh!]
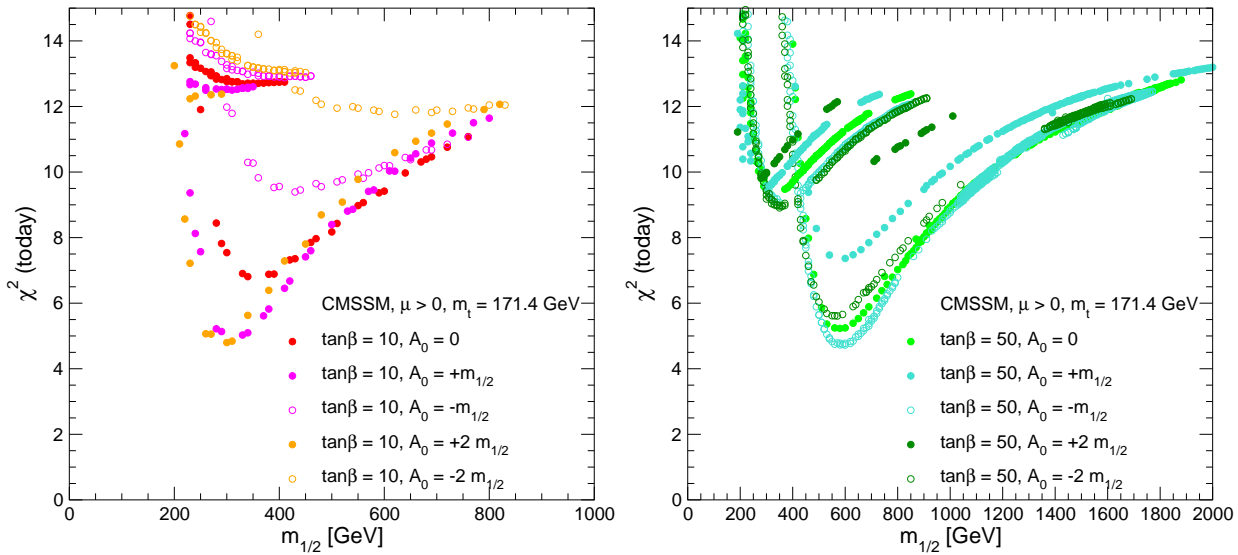

\begin{center}
\includegraphics[width=.48\textwidth]{ehow5.CHI11a.1714.cl.eps}
\includegraphics[width=.48\textwidth]{ehow5.CHI11b.1714.cl.eps}
\caption{%
The combined $\chi^2$~function for electroweak precision
observables and $B$-physics observables, 
evaluated in the CMSSM for $\tan \beta = 10$ (left) and
$\tan \beta = 50$ (right) for various discrete values of $A_0$. We use
$m_t = 171.4 \pm 2.1$~GeV and $m_b(m_b) = 4.25 \pm 0.11$~GeV, and 
$m_0$ is chosen to yield the central value of the cold dark matter
density indicated by WMAP and other observations for the central values
of $m_t$ and $m_b(m_b)$~\protect\cite{EHOWW}.}
\label{fig:fit}
\end{center}
\vspace{-1em}
\end{figure*}

Is there any preference for any particular range of sparticle masses within
this allowed band? Fig.~\ref{fig:fit} shows the $\chi^2$ distributions for
global fits to precision electroweak and $B$-decay data, assuming
$\tan \beta = 10$ (left) and $\tan \beta = 50$ (right)~\cite{EHOWW}. We see that relatively
low values of $m_{1/2} \sim 300, 600$~GeV are favoured, essentially by
$g_\mu -2$ (though there is also some support from $m_W$). Results
from a more complete analysis of parameter space using a larger set of
observables (with better graphics) are given in~\cite{Tools}. There may
be good reason to hope that supersymmetry might be detectable
at the LHC with 1~fb$^{-1}$ of integrated luminosity.

\section{Looking for Supersymmetry at the LHC}
\label{LHC}

The classic supersymmetric signature is missing transverse energy
carried away by dark matter particles. The Tevatron collider has already provided
important limits on gluinos and squarks: $m_{\tilde g} > 290$ to 410~GeV and 
$m_{\tilde q} > 375$~GeV~\cite{Duperrin,Shamim} and has also provided important upper limits on
trilepton final states as might arise from chargino and neutralino production~\cite{Mundal}.
Even with low initial luminosity, the LHC will immediately have sensitivity to
gluino and squark masses far beyond the Tevatron limits. However, the
missing-energy search will not be without backgrounds~\cite{Mangano}, it will be necessary
to understand very well the ATLAS and CMS detectors.

A possible strategy for classic supersymmetry sear- ches (and discovery?) at 
the LHC is~\cite{Spiropoulou}:
(i) search for single lepton + missing-energy events,
(ii) search for all combinations of dilepton + missing-energy events,
(iii) search for trilepton + jet events,
(iv) search for ${\bar b} b$ + lepton events,
(v) search for zero-lepton + missing-energy events,
etc. In addition, there will be searches for the photons characteristic of
gauge-mediated scenarios and the metastable particles that might appear in
scenarios with a gravitino LSP. Fig.~\ref{fig:POFPA} shows the sensitivity of the LHC
to the gluino mass (both for five-$\sigma$ discovery and 95~\% exclusion)
as a function of the integrated luminosity. For example, with 1~fb$^{-1}$ the LHC
might be able to discover a gluino weighing up to 1.7~TeV, or exclude a gluino
weighing less than 2.1~TeV.

\section{The LHC Reach and Linear Colliders}
\label{LHCLC}

The results of the LHC search for gluinos will carry important implications
for future linear colliders~\cite{POFPA}, as illustrated in Fig.~\ref{fig:POFPA}. 
Specifically, concentrating on the 
coannihilation region and models with unification of gaugino
masses at the GUT scale, such as the CMSSM, discovery with 1~fb$^{-1}$
would suggest that the $e^+ e^- \to \chi \chi$ threshold lies below 650 GeV,
whereas exclusion would exclude a threshold below 800~GeV. Well beyond
the initial LHC luminosity, discovery of the gluino at the SLHC with a `year'
at a luminosity of $10^{35}$~cm$^{-2}$s$^{-1}$ would tell us that the 
$e^+ e^- \to \chi \chi$ threshold lies below about 1.3~TeV. The LHC will
tell us what energy a linear collider will need to study supersymmetry.

As is well known, the ILC would be able to make accurate measurements of 
the masses and couplings of any sparticles within its kinematic reach~\cite{Choi}, and
these measurements would have invaluable synergies with LHC
measurements, for example by probing models of supersymmetry breaking
by testing unification ideas~\cite{Choi}. However, supersymmetrists should be aware
that not all scenarios with large cross sections at the LHC will necessarily
be observable at the ILC~\cite{Lillie}. A preliminary study of 242  such scenarios found that 158 or
65~\% have no observable signal at the ILC with 500~GeV, and further
investigation shows that the unobservable percentage actually rises to 75~\%.

\section{Supersymmetric Higgs Bosons}
\label{MSSMH}

The LHC also has good prospects of discovering supersymmetric Higgs bosons,
being able to cover entire generic $(m_A, \tan \beta)$ planes at least
(but perhaps only) once. However, most points in the $(m_A, \tan \beta)$ planes
corresponding to fixed values of $\mu, m_{1/2}$ and $m_0$ do not have a cold
dark matter density within the range favoured by WMAP and other astrophysical
and cosmological measurements~\cite{Olive}.

The situation changes in models with non-universal scalar masses. Upper
limits on flavour-changing neutral interactions disfavour non-universal
masses for sfer- mions in different generations but the same quantum numbers,
e.g., $d_R, s_R$ and $b_R$. Also, GUT models favour universal scalar masses for
squarks and sleptons in the same multiplets, e.g., $d_L, u_L, u_R$ and $e_R$
($d_R$ and $e_L$) in the {\bf 10} (${\mathbf {\bar 5}}$) of SU(5), or all the 
fermions of the same generation in SO(10). However, there is no known
reason why the supersymmetry-breaking contributions to Higgs scalar masses
should not be non-universal, a scenario known as the NUHM.

\begin{figure*}[htb!]
\vspace{10mm}
\begin{center} 
\includegraphics[width=.49\textwidth]{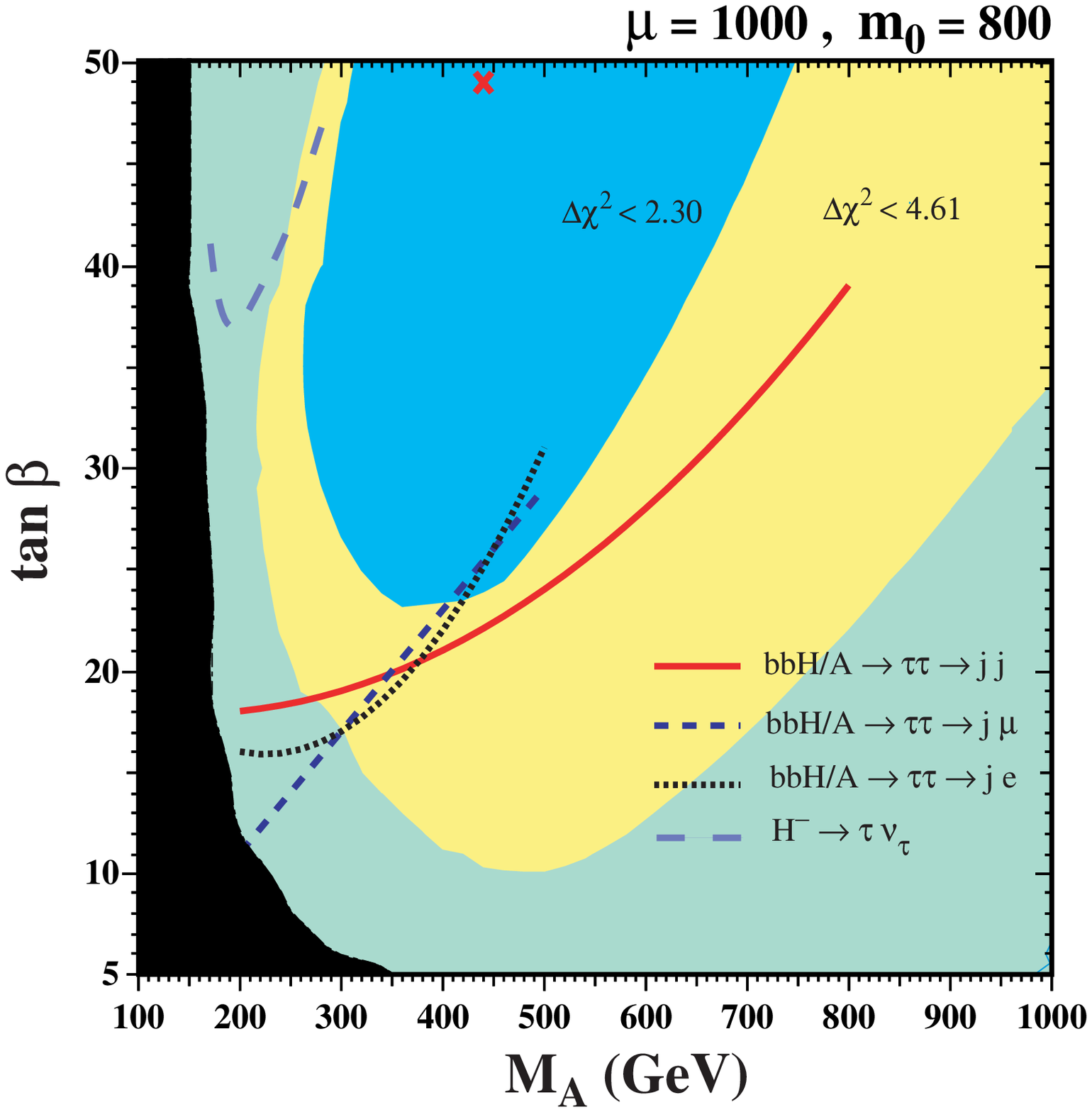}
\includegraphics[width=.49\textwidth]{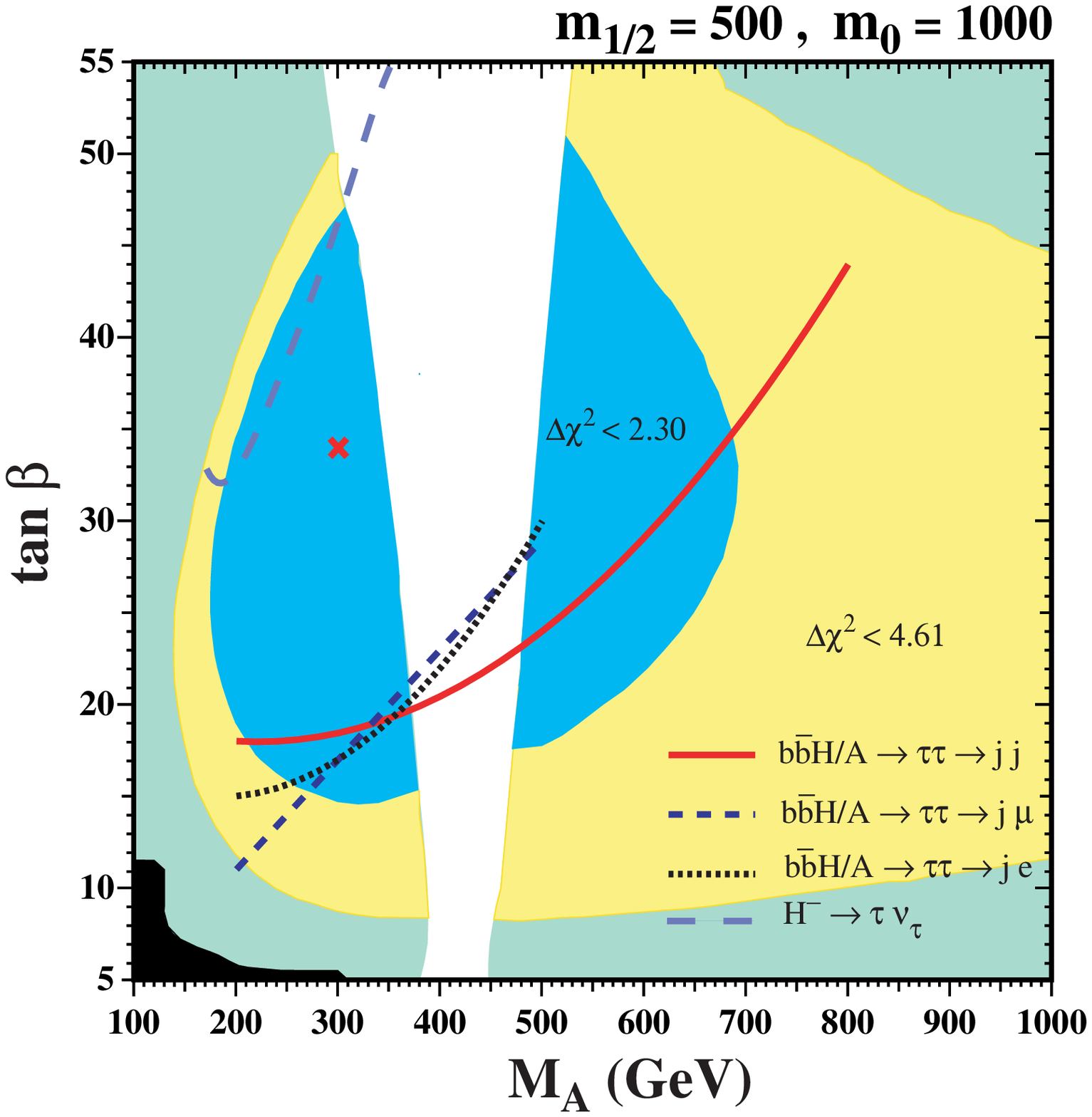}
\includegraphics[width=.49\textwidth]{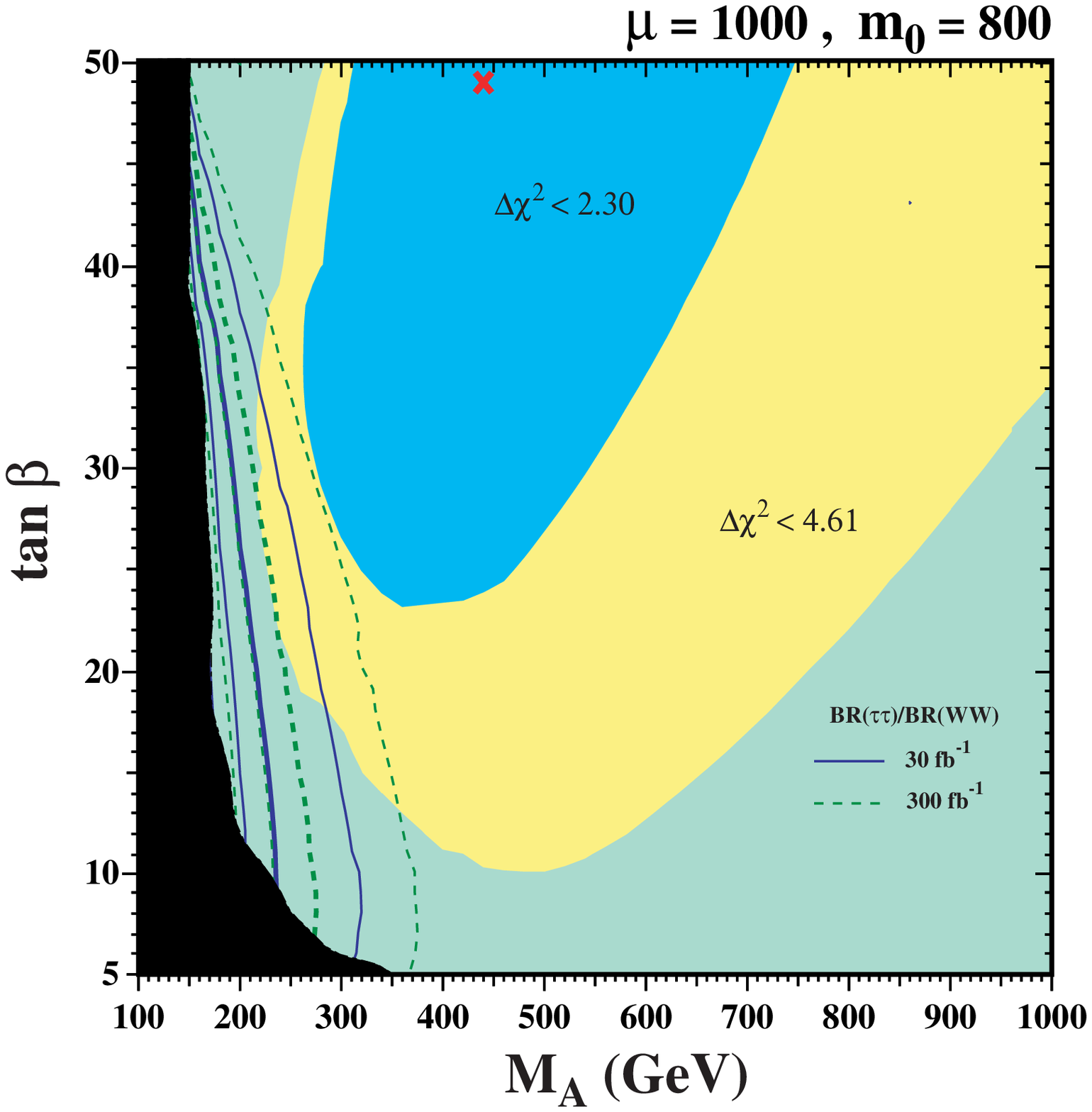}
\includegraphics[width=.49\textwidth]{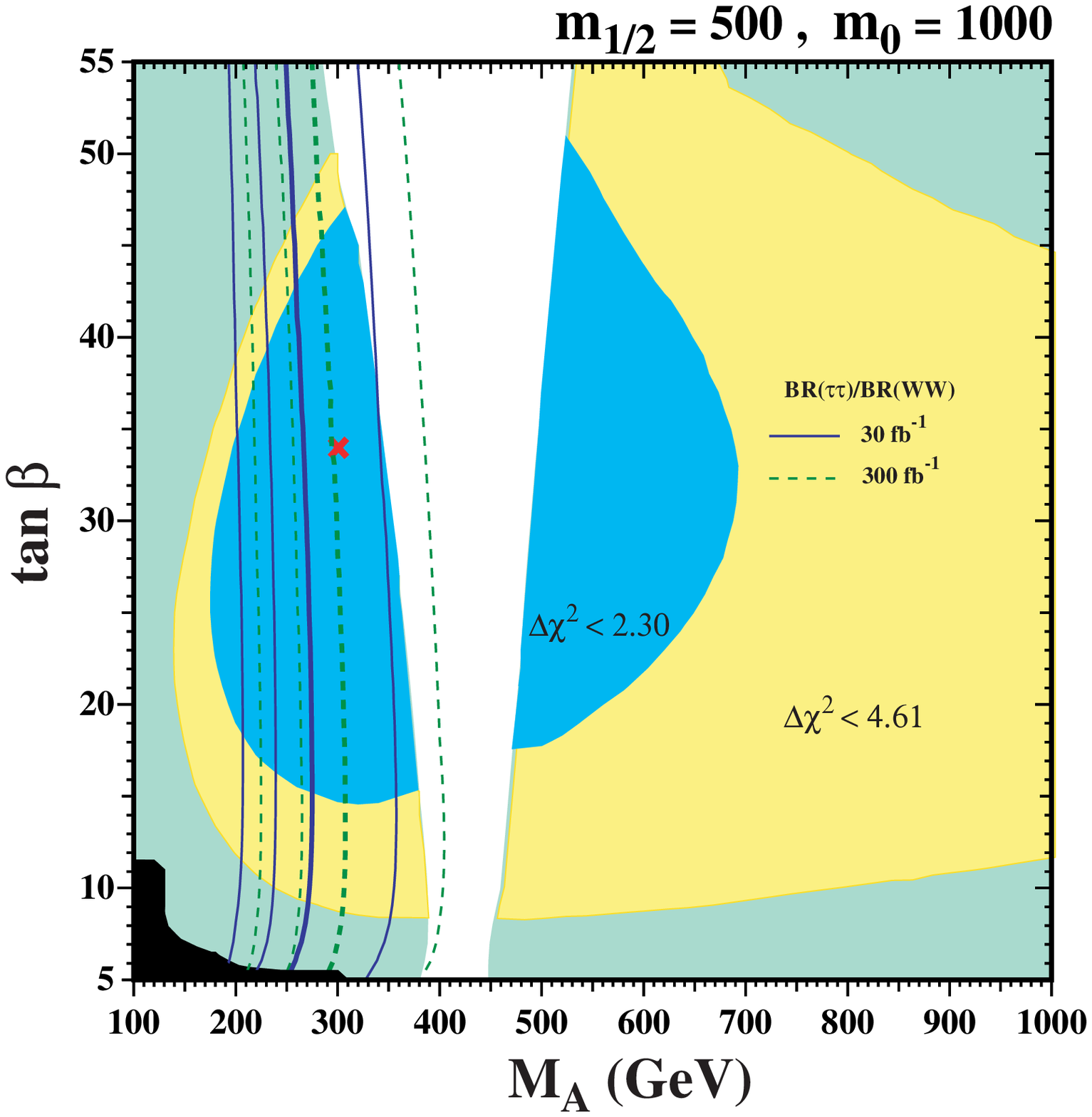}
\caption{%
WMAP-compatible $(M_A, \tan \beta)$ planes for two NUHM benchmark surfaces, displaying 
(top) the 5-$\sigma$ discovery contours 
for $H/A \to \tau^+ \tau^-$ at the LHC with 60~or 30~fb$^{-1}$
(depending on the $\tau$ decay channels) and for $H^\pm \to \tau^\pm \nu$
detection in the
CMS detector when $M_{H^\pm} > m_t$, and (bottom)
the 1-, 2-, 3- and 5-$\sigma$ contours (2-$\sigma$ in bold) 
for SUSY-induced deviations from the SM value for the ratio
$BR(h \to \tau^+\tau^-)/BR(h \to WW^*)$ at the LHC with 30~or 
300~fb$^{-1}$~\protect\cite{EHHOW}.}
\label{fig:EHHOW}
\end{center}
\vspace{1em}
\end{figure*}

We have recently proposed and studied four WMAP-compatible $(m_A, \tan\beta)$
surfaces in the NUHM, using the non-universality parameters of the two Higgs
multiplets to keep $\Omega_\chi h^2$ within the range favoured by WMAP 
et al~\cite{EHHOW,Olive}.
One of these WMAP-compatible surfaces have fixed $m_0 = 800$~GeV, 
fixed $\mu = 1000$~GeV, and varying $m_{1/2}  \sim 9/8 m_A$, and another has
fixed $m_{1/2} = 500$~GeV, fixed $m_0 = 1000$~GeV and varying $\mu \sim 250$ 
to 400~GeV. In each of these planes, we have made global fits to the
electroweak precision and $B$ observables, and analyze the models'
detectability at the Tevatron, LHC and ILC. For example, Fig.~\ref{fig:EHHOW}
displays the detectability of the $H/A \to \tau \tau$ signals at the Tevatron and LHC,
and the accuracy with which the ILC could measure the $h \to {\bar b}b$ and $WW$
branching ratios.

\section{Gravitino Dark Matter?}
\label{GDM}

\begin{figure}[h]
\includegraphics[height=3in]{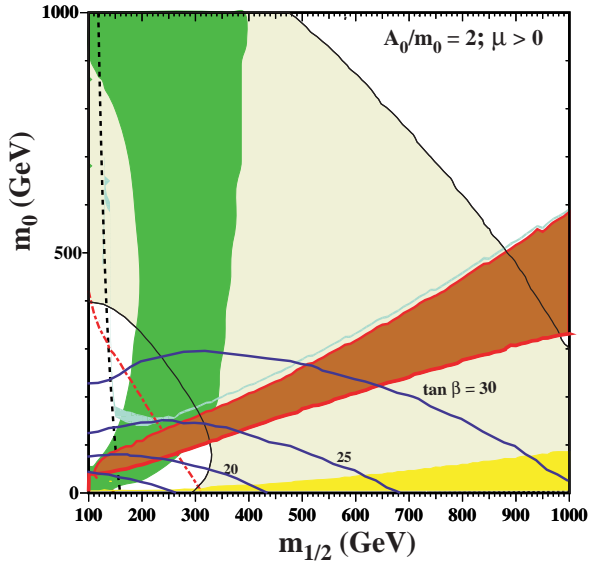}
\caption{\label{fig:GDM}
{Example of an mSUGRA $(m_{1/2}, m_0)$ plane with contours of $\tan \beta$ 
superposed, for $\mu > 0$ and $A_0/m_0 = 2.0, B_0 =
A_0 -m_0$~\protect\cite{Olive}. The regions excluded by 
LEP are indicated, as are excluded by $b
\to s \gamma$ decay (medium green shading), and the region 
favoured by $g_\mu - 2$ is light (beige) shaded. The region favoured 
by WMAP in the neutralino LSP case has light (blue)
shading, and the regions with a stau NLSP that would be
allowed by the cosmological BBN constraint (neglecting bound-state
effects~\protect\cite{Pospelov}) is light (yellow) shaded.}}
\end{figure}

As already mentioned, it is possible that the the LSP might be the gravitino~\cite{Steffen},
which would therefore provide the dark matter (GDM), and this is a
generic possibility even in minimal supergravity (mSUGRA), as shown in
Fig.~\ref{fig:GDM}. After taking into account the LEP and $b \to s \gamma$
constraints, there is a (pale blue) strip where the lightest neutralino is the LSP,
a disallowed (brown) wedge where the LSP would be the lighter stau ${\tilde \tau_1}$~\cite{Steffen},
and another (yellow) wedge where the ${\tilde \tau_1}$ is the next-to-lightest
sparticle (NLSP) and metastable, but its decays do not upset the cosmological
light-element abundances.

It is a feature of any GDM scenario with gravity-mediated supersymmetry breaking that
the NLSP has a very long lifetime due to gravitational decay, which might be
measurable in hours, days, weeks, months or even years. Generic possibilities
in models with non-universal scalar masses include the lightest neutralino $\chi$
as well as the lightest charged slepton (probably the ${\tilde \tau_1}$)~\cite{ERO}, 
a sneutrino~\cite{Covi}, 
or even the lighter stop quark ${\tilde t_1}$~\cite{Diaz}.

\begin{figure*}
\resizebox{2.0\columnwidth}{!}
{\includegraphics{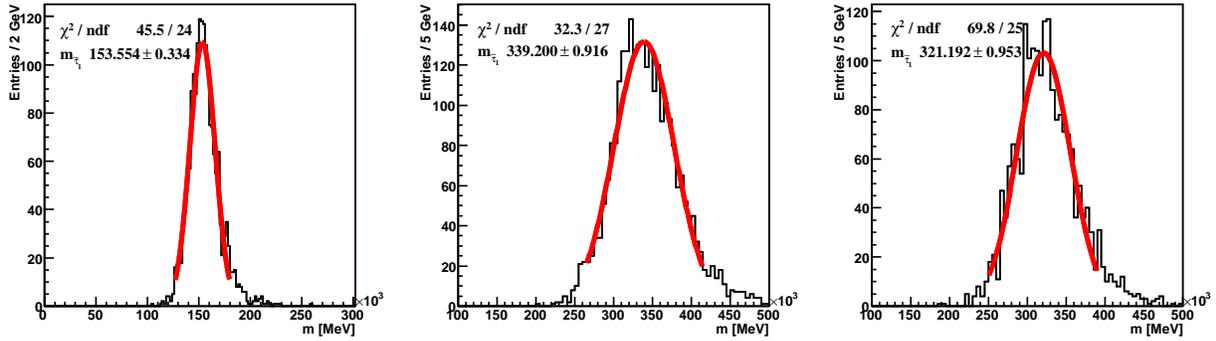}}
\caption{\label{fig:ERO2}The mass of a metastable stau could be measured
quite accurately at the LHC~\protect\cite{ERO2}, as exemplified in three benchmark 
scenarios~\protect\cite{Bench3}. }
\end{figure*}

A study of the ATLAS sensitivity to a metastable ${\tilde \tau_1}$ shows that events
containing them would be selected with high efficiency by several separate triggers, 
via combinations of energetic muons, electrons or jets, without looking for a
candidate ${\tilde \tau_1}$ track~\cite{ERO2}. These could then be selected with high
efficiency in an off-line analysis, and the ${\tilde \tau_1}$ mass determined by a
combination of momentum and time-of-flight measurements, as shown in
Fig.~\ref{fig:ERO2}. The detected
${\tilde \tau_1}$'s may then be combined with jets and leptons in the final states
to reconstruct invariant-mass peaks for heavier sparticles in such a GDM scenario~\cite{ERO}.

The ${\tilde \tau_1}$'s produced at the LHC would typically be only moderately
non-relativistic, with $\beta \gamma \sim 2$. The ${\tilde \tau_1}$'s with
$\beta \gamma < 1$ would leave the central tracker 	after the next beam crossing,
those with $\beta \gamma < 1/2$ would be stopped by $\sim 10$~m of matter,
and those with $\beta \gamma < 1/4$ would likely be trapped inside the
ATLAS calorimeter. We have wondered whether those with $1/4 < \beta \gamma < 1/2$
could be dug out of the LHC experimental cavern walls~\cite{Bench3}. The idea would be to use 
the muon system to locate the impact point on the cavern wall with an uncertainty 
$< 1$~cm and the impact angle with an accuracy $\sim 10^{-3}$~radians. One might
then bore into the cavern wall and remove a core from the rock where the ${\tilde \tau_1}$
should have stopped, then wait for it to decay and observe its decay products.
However, one must beware of the radioactivity induced by LHC collisions. Once a month
it is planned to make a technical stop of the LHC for two days, during which one could
work in the cavern. This would be useful if the ${\tilde \tau_1}$ lifetime is more than about
$10^6$~s, but not if it is much shorter.

There is, however, an additional cosmological effect that needs to be taken into account.
If the NLSP is charged (like the ${\tilde \tau_1}$), it may form bound states in the early
Universe, which would have additional effects on the light-element abundances~\cite{Pospelov}.
Under certain circumstances, these might even {\it improve } the agreement of
theoretical calculations with the observed abundances of $^{6,7}$Li~\cite{CEFOS}. However, this is
possible only in very restricted regions of the GDM parameter space, in which the
NLSP is relatively heavy (at least in the first examples studied).

\section{Complexification of the CMSSM}

\begin{figure}
\includegraphics[width=0.55\textwidth,angle=0]{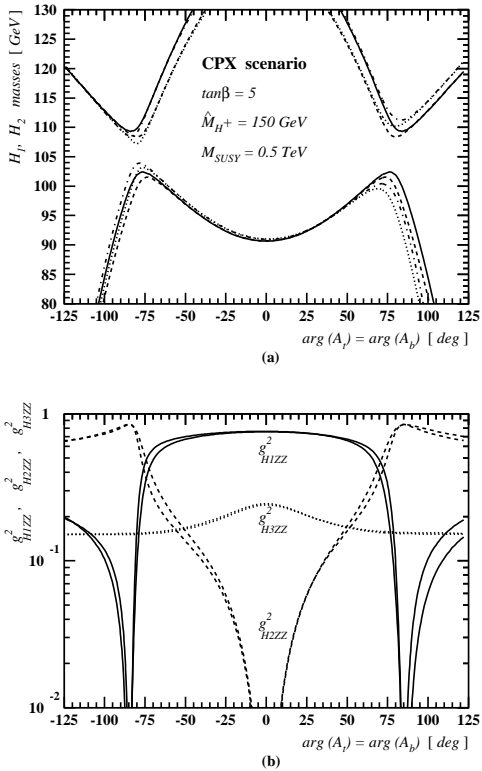}
\caption{Numerical estimates of (a) the $H_{1,2}$-
effective-potential and pole masses and (b)~$g^2_{H_iZZ}$ as functions
of Arg$(A_0\mu)$, in a CP-violating scenario with $M_{\rm SUSY} =
0.5$~TeV, Arg $(m_{1/2}\mu) = 0$ and $90^\circ$. In
plot~(a), the effective-potential mass $M_{H_1}~(M_{H_2})$ is
indicated by a solid (dash-dotted) line for Arg $(m_{1/2}\mu) = 0~(90^\circ )$, 
and its pole mass
$\widehat{M}_{H_1}~(\widehat{M}_{H_2})$ by a dashed (dotted) line for
Arg $(m_{1/2}\mu) = 0~(90^\circ )$~\protect\cite{CEPW}.}
\label{fig:CPX}       
\end{figure}

Assuming universal soft supersymmetry-breaking parameters, as in the CMSSM,
there are two new CP-violating phases beyond those in the SM, namely
Arg $(m_{1/2}\mu)$ and Arg$(A_0\mu)$~\cite{Kraml}. At the loop level, these induce mixing
between the CP-even and -odd neutral Higgs bosons of the CMSSM, so that
$(h, H, A) \to (H_1, H_2, H_3)$ with indefinite CP. As seen in Fig.~\ref{fig:CPX},
the new CP-violating
parameters affect the couplings of the MSSM Higgs bosons as well as their
masses. The phases could in principle allow the lightest CMSSM Higgs boson to
be lighter than in the usual limit in the MSSM with real parameters, but they are
subject to important constraints imposed by upper limits on electric dipole
moments. There are prospects for probing these phases at the LHC, in both
the Higgs and sparticle sectors.

\section{Supersymmetric Flavour Physics}

\begin{figure}
\includegraphics[width=0.45\textwidth,angle=0]{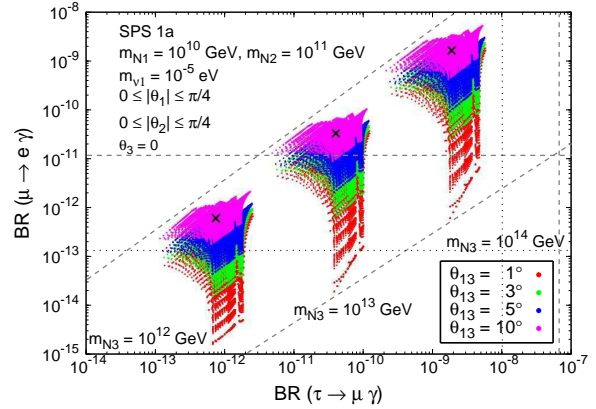}
\caption{Correlation between $BR(\mu \to e\,\gamma)$ and 
      $BR(\tau \to \mu\,\gamma)$ for different $m_{N_3}$, displaying the impact of $\theta_{13}$
      with a scan over $\theta_i$~\protect\cite{Herrero}. The horizontal and vertical 
      dashed (dotted) lines denote the experimental bounds (future
      sensitivities).}
\label{fig:Herrero}       
\end{figure}

The flavour and CP structure of any new physics at the TeV scale is tightly
constrained by the continuing agreement of data from the $B$ factories and
the Tevatron with the predictions of the SM, e.g., their measurements of
$b \to s \gamma$ and $B_u \to \tau \nu$, and their upper limits on
$B_s \to mu^+ \mu^-$. Improvements in these measurements and limits are places
to look for supersymmetric flavour physics, and other opportunities may be
provided by $K$ physics~\cite{Smith}, for example in the search for violations of flavour
universality in $K \to e \nu$ and $K \to \mu \nu$ 
decays~\cite{Petronzio}, or in $K \to \pi \nu \nu$ decays.
Charged leptons may also play roles in unravelling the supersymmetric flavour
puzzle. For example, in supersymmetric extensions of the see-saw model for
neutrino masses, the lepton-flavour-violating processes $\mu \to e \gamma$
and $\tau \to \mu (e) \gamma$ may occur at observable rates~\cite{Masiero,Herrero,Deppisch},
as illustrated in Fig.~\ref{fig:Herrero}.

\section{Suggestions from String Theory?}

The full enormity of the ambiguity in the string vacuum has sunk in only recently,
with numbers ${\cal O}(10^{500})$ being bandied about~\cite{Linde}. This ambiguity arises
because there are certainly millions and perhaps billions of consistent
compactifications of strings on manifolds in extra dimensions, and each of these
has dozens or hundreds of topological cycles through which there may be
topological fluxes taking any of dozens of values. Somewhere in this landscape of an
enormous number of string vacua, it is suggested there may be one with
a vacuum energy in the range indicated by the cosmology dark energy.
The question then arises how the Universe chooses which of these vacua.
One may also wonder whether, since nature apparently has the opportunity
to choose a small vacuum energy, perhaps it also chooses a small value
of $m_W$, in which case there might be no need for supersymmetry to render the 
choice natural.

\begin{figure}
\includegraphics[width=0.45\textwidth,angle=0]{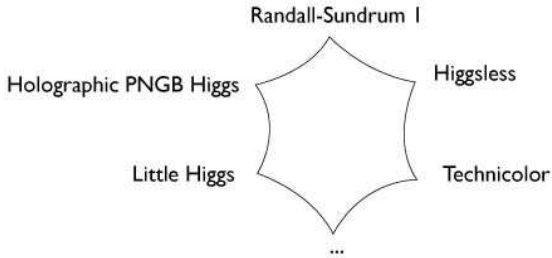}
\caption{An illustration of the space of possible alternatives to supersymmetry
at low energies~\protect\cite{Cheng}.}
\label{fig:Cheng}       
\end{figure}

Indeed, ideas for models without supersymmetry 
(or even a Higgs boson) were also discussed here~\cite{Cheng},
and a unified discussion of alternatives has been presented. As illustrated
in Fig.~\ref{fig:Cheng}, there is a continuum
of alternatives to the supersymmetric paradigm, ranging from little Higgs models to
holographic pseudo-Nambu-Goldstone-boson Higgs models to Randall-Sun- drum
scenarios to Higgsless models to technicolour models and back again. The good
news is that many (most? all?) of these scenarios can be tested at the LHC.

An different question revived by the string landscape within the
supersymmetric paradigm is whether we live in a
metastable vacuum~\cite{Seiberg}, a possibility discussed a long time ago as an exotic
possibility in the framework of global supersymmetry~\cite{ELR}. If there are indeed myriads
of consistent string vacua, it seems difficult to see why we should be living in the
one that is energetically preferred.

\begin{figure}[ht!]
\includegraphics[height=2.9in]{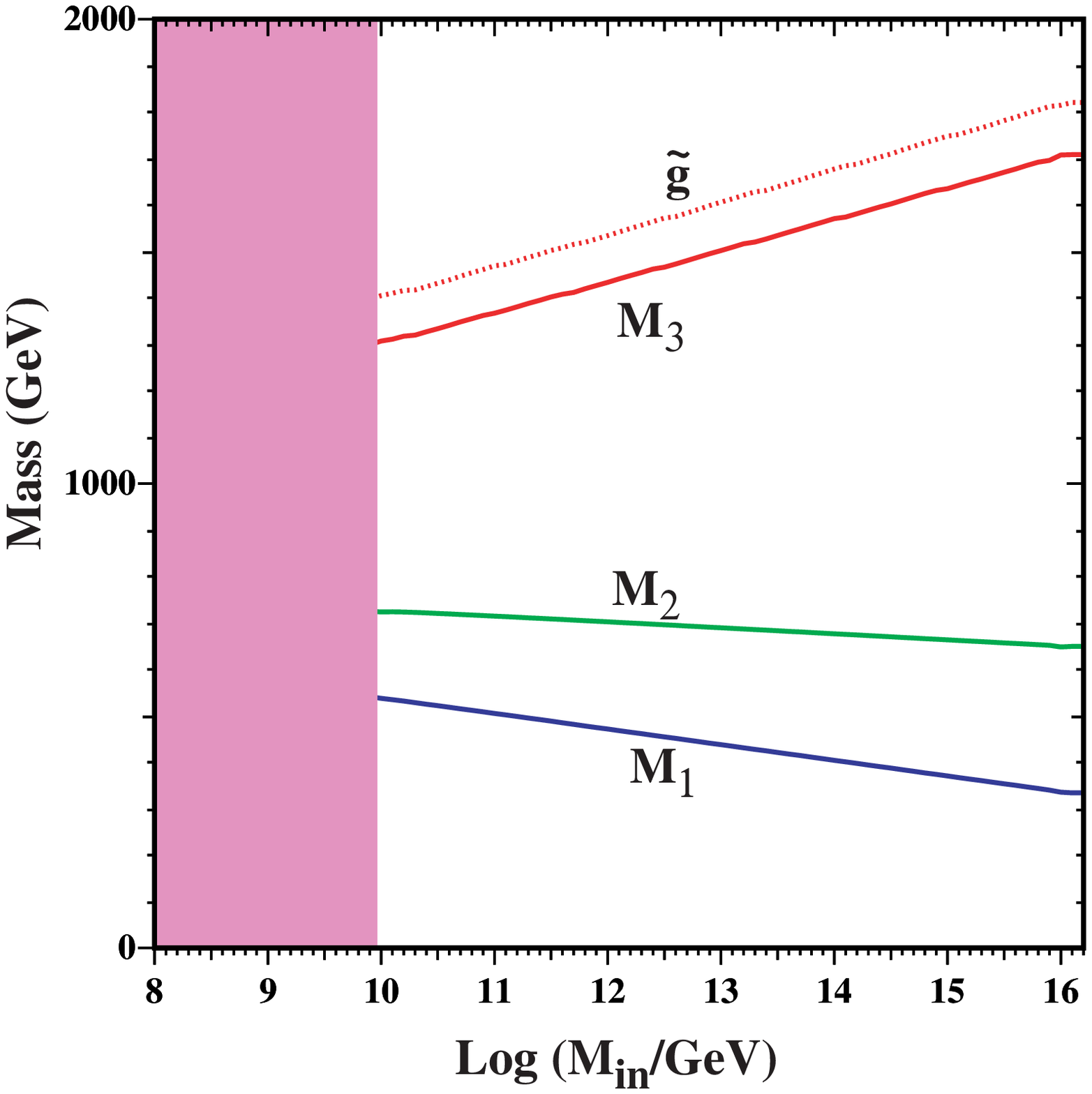}
\includegraphics[height=3in]{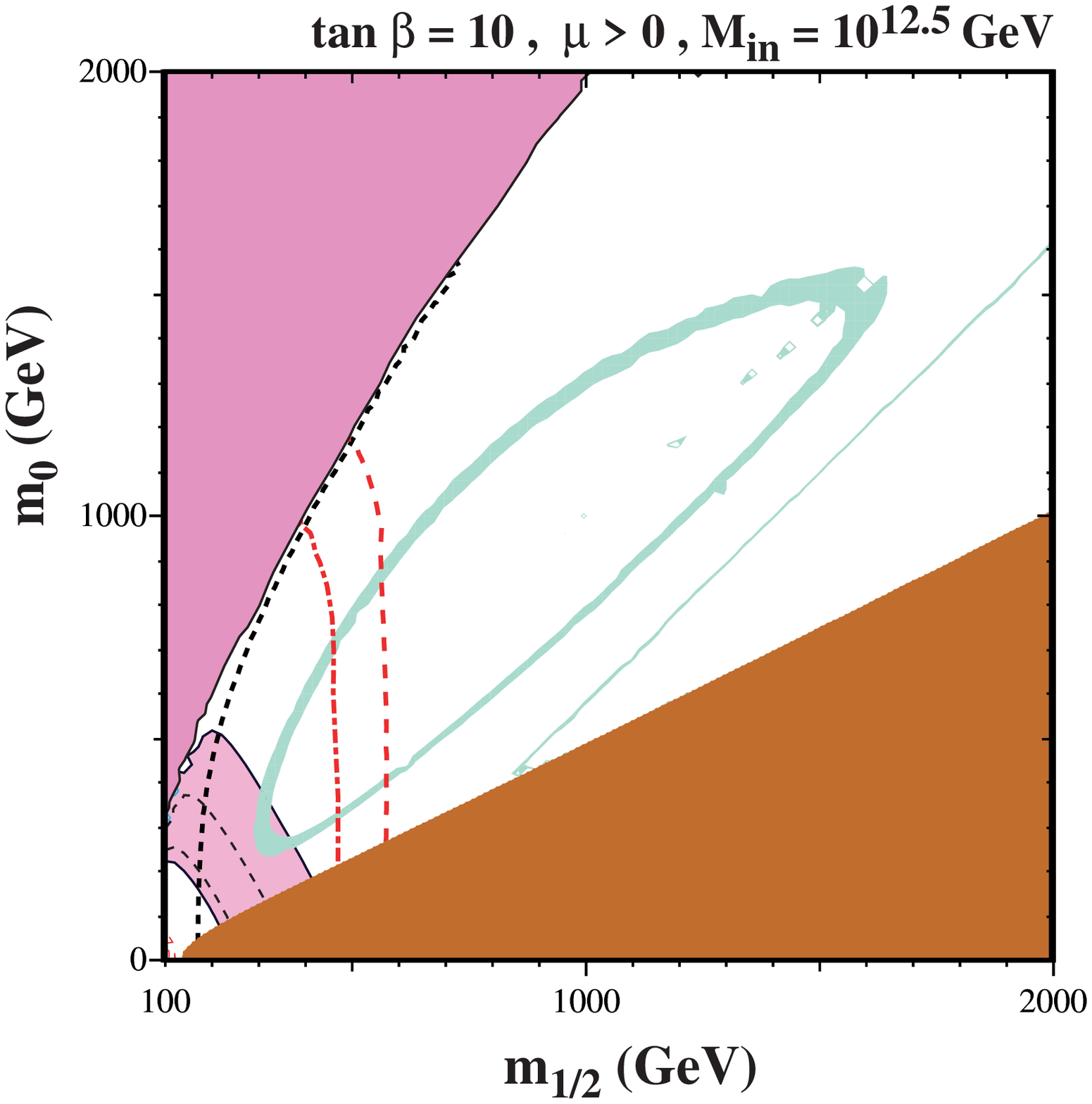}
\caption{\label{fig:Pearl}
{Evolution (top) of the gaugino mass parameters and the physical gluino mass
as the effective `mirage' unification scale $M_{in}$ is reduced, and (below)
an example of an $(m_{1/2}, m_0)$ plane with $\tan \beta = 10$ and 
$A_0 = 0$ and  $M_{in} = 10^{12.5}$ GeV, using the same notation as in
Fig.~\protect\ref{fig:CMSSM}. The region favoured 
by WMAP is very different from that in the CMSSM with GUT-scale universality. }}
\end{figure}

The mainstream hope would be that string theory would add value to the
MSSM by predicting the exact spectrum and grand unification, incorporating
gauge and/or Yukawa unification and the see-saw mechanism and providing an
explicit mechanism for breaking supersymmetry, e.g., via gaugino condensation~\cite{Nilles}.
One interesting variant of the conventional CMSSM scenario is the possibility
of `mirage unification', in which gaugino masses unify below the GUT scale
as a result of mixed modulus and anomaly contributions to gaugino masses~\cite{mirage}:
\begin{equation}
M_a \; = \; M_s ( \rho + b_a g_a^2).
\label{mirage}
\end{equation}
Lowering the unification scale could have a dramatic effect on the phenomenology
discussed previously in the CMSSM context. As seen in the top panel of Fig.~\ref{fig:Pearl}, 
the expected values of the sparticle masses change as the effective universality
scale $M_{in}$ is reduced and, as a consequence,
the regions of parameter space favoured by cosmology may change 
significantly~\cite{Sandick,Olive},
as seen in the lower panel of Fig.~\ref{fig:Pearl}.
The LHC may soon tell us, in more ways than one, whether supersymmetry is a
mirage!

\section{Conclusions}

LEP and the Tevatron have already advanced in the quests for supersymmetry
and the Higgs boson, and these searches being continued by the Tevatron. In parallel,
searches for supersymmetry have been underway in low-energy precision
physics and in direct and indirect searches for dark matter, and will continue
during the LHC era. However, the LHC will be the first accelerator to reveal to us
directly what new physics exists at the electroweak scale. Without results from
the LHC, starting around 2010, we will not know what major 
subsequent new accelerator investments would be optimal. One possibility, 
that would optimize the scientific return from immense
investment that the community has made in the LHC, would be to improve its
luminosity, perhaps by an order of magnitude. This would surely be interesting in
many supersymmetric and other scenarios for physics beyond the SM. On a
longer time-scale, there is general agreement that a linear $e^+ e^-$
collider would be an ideal tool for studying in detail any new physics revealed
by the LHC, {\it provided} it lies within the accessible energy range. Only time
and the LHC will tell us whether the ILC will have sufficient energy, or whether
physics will demand higher energy, as could be provided by CLIC.

\end{document}